\documentclass[final,5p,times,twocolumn,sort&compress]{elsarticle}

\usepackage{amssymb}
\usepackage{amsmath}
\usepackage{caption}
\usepackage{subcaption}
\usepackage{booktabs}
\newcommand{\Fig}[1]{Fig.~\ref{#1}}

\journal{Physics Letters B}
\usepackage{eso-pic}
\AddToShipoutPicture*{%
  \AtPageLowerLeft{%
    \ifnum\value{page}=1
    \hspace{15mm}\raisebox{15mm}
    \fi
  }
}

\begin{document}

\begin{frontmatter}



\title{Excitation function of femtoscopic Lévy source parameters of pion pairs in EPOS4}


\author[ELTE]{Yan Huang}
\ead{huang@ttk.elte.hu}
\author[ELTE]{Mátyás Molnár}
\author[ELTE]{Dániel Kincses}
\author[ELTE]{Máté Csanád}

\affiliation[ELTE]{organization={Department of Atomic Physics, ELTE Eötvös Loránd University},
            addressline={Pázmány Péter sétány 1/A}, 
            city={Budapest},
            postcode={H-1117}, 
            state={},
            country={Hungary}}

\begin{abstract}
Three-dimensional (3D) femtoscopic source parameters of pions provide a sensitive probe of the space-time structure of particle-emitting sources in high-energy heavy-ion collisions. Compared to one-dimensional measurements, three-dimensional femtoscopy not only provides a valuable cross-check but also offers a more complete characterization of the source geometry and its dynamical evolution. Particularly, differences between the ``out'' and ``side'' directions are sensitive to signals of a strong first-order phase transition, while the collision-energy dependence of Lévy radii may reveal non-monotonic features related to the equation of state. 

In this work, we systematically investigate the transverse mass ($m_T$) and collision-energy ($\sqrt{s_{NN}}$ ) dependence of the three-dimensional femtoscopic parameters of pion pairs with Lévy-type sources in the STAR Beam Energy Scan (BES) range from $\sqrt{s_{NN}} = 7.7$ to $200$ GeV using the EPOS4 model. The analyzed parameters include the Lévy index $\alpha$, the correlation strength $\lambda$, and the three-dimensional radii $R_{\mathrm{out}}$,  $R_{\mathrm{side}}$ and $R_{\mathrm{long}}$, corresponding to the outward, sideward, and longitudinal (beam) directions. Derived quantities such as the radius difference $R_\mathrm{diff} = R_\mathrm{out}^2 - R_\mathrm{side}^2$ and the ratio $R_\mathrm{out}/R_\mathrm{side}$ are also investigated. The results show that the extracted radii $R_{\mathrm{side}}$ and $R_{\mathrm{long}}$ decrease with increasing transverse mass and increase gradually with collision energy, while $R_{\mathrm{out}}$ shows little energy dependence. The Lévy index $\alpha$ exhibits only a mild dependence on $m_T$ and collision energy, whereas the correlation strength $\lambda$ shows a clear $m_T$ dependence and generally decreases with increasing collision energy.
A comparison with EPOS3 results indicates general agreement within approximately $2\sigma$, with the notable exception of $R_\mathrm{side}$, which is systematically smaller in EPOS4.
\end{abstract}




\end{frontmatter}




\section{Introduction}
\label{introduction}

High-energy heavy-ion collisions provide a unique environment for studying strongly interacting matter under extreme conditions of temperature and density. One of the most sensitive tools for probing the space-time structure of the particle-emitting source is femtoscopy, which relies on the measurements of two-particle momentum correlations at small relative momenta.

Traditionally, the source distribution was modeled by a Gaussian form. However, in recent years, various experiments--from SPS~\cite{Adhikary2023SPS, Porfy2024aSPS, Porfy2024bSPS}, through RHIC~\cite{Adare2018RHIC, Kovacs2023RHIC, Mukherjee2023RHIC, Abdulameer2024RHIC, Kincses2024RHIC}, to LHC~\cite{Tumasyan2024LHC, Korodi2023LHC}--have shown that Lévy-stable source distributions provide a better description of the measured correlation functions, particularly in cases where anomalous diffusion or long tails appear in the emission profile. Several physics mechanisms in high-energy heavy-ion collisions may contribute to such non-Gaussian behavior, including critical phenomena, jet fragmentation, and anomalous diffusion associated with the long-range transport in the expanding medium~\cite{Csorgo2005Levy, Csanad2007Levy, METZLER20001, universe10020054}. A three-dimensional (3D) Lévy femtoscopy method has been recently developed and applied to investigate pion sources in relativistic heavy-ion collisions~\cite{Kincses:2024lnv}. This study provided a comprehensive framework for extracting the Lévy index of stability, source radii, and correlation strength, offering improved sensitivity to non-Gaussian features of the source.

A particularly intriguing aspect of Lévy femtoscopy is its potential to reveal signatures of the QCD critical point. Near the critical point, long-range correlations are expected, leading to a non-trivial, possibly non-monotonic energy dependence of the Lévy index $\alpha$~\cite{Csanád_2025}. Motivated by this, we systematically analyze the energy dependence of femtoscopic source parameters within the EPOS4 event generator framework. This provides a crucial non-critical theoretical baseline for interpreting future experimental data and for exploring possible signs of critical behavior in particle correlations.

\section{Methods}
Our analysis is based on the latest 4.0.3 version of the general-purpose Monte Carlo event generator EPOS, to simulate heavy-ion collisions at different collision energies~\cite{PhysRevC.108.064903, PhysRevC.111.014903}. 
Compared to EPOS3, EPOS4 implements a fully self-consistent, energy-momentum conserving parallel scattering scheme with subscattering dependent saturation scales. This represents a genuine paradigm shift that leads to harder high-multiplicity events and consequently modifies the collective source characteristics.

\subsection{Distance distribution measurement and extraction of key parameters}
\begin{figure*}
  \centering
  \includegraphics[width=0.95\linewidth]{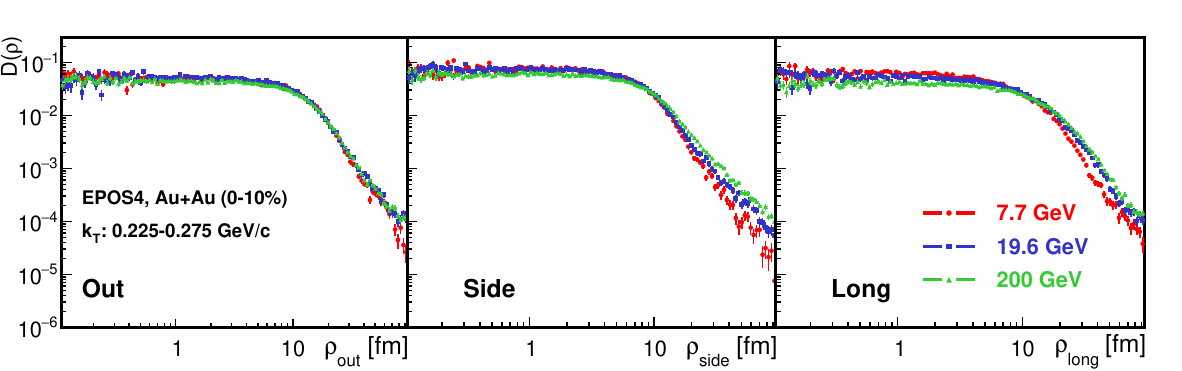}
  \caption{Pair distance distribution projections $D(\rho_\mathrm{out}^\mathrm{LCMS})$, $D(\rho_\mathrm{side}^\mathrm{LCMS})$, and $D(\rho_\mathrm{long}^\mathrm{LCMS})$ from EPOS4 simulation of 0-10\% Au+Au collisions for the $k_T$ bin $0.225-0.275~\rm{GeV}/c$ at 7.7, 19.6 GeV, and 200 GeV collision energies. }
  \label{fig:D_rho}
\end{figure*}

In simulations, the pair distance distribution $D(\vec{\rho})$ can be directly obtained, where $\vec{\rho}$ represents the three-dimensional spatial separation vector of the particle pairs. And representative examples of the distributions are shown in \Fig{fig:D_rho}, which are discussed in detail in Section 3.2. This allows the extraction of key source parameters such as the Lévy index \(\alpha\), the source size \(R\), and the correlation strength \(\lambda\). In experimental measurements, however, the spatial distributions are not directly accessible. Instead, one measures the correlation function in momentum space (as a function of momentum difference $\vec q$), which encodes information about both the spatial distribution $D(\vec{\rho})$ and the pair wave function $\psi_{\vec{q}}(\vec{\rho})$:
\begin{equation}
C(\vec{q}) = \int D(\vec{\rho}) \, \bigl|\psi_{\vec{q}}(\vec{\rho})\bigr|^2 \, d^3 \vec{\rho}.
\end{equation}
For a Lévy-stable single-particle source, the pair distance distribution can be expressed as ~\cite{Nolan2013LevyEq}:
\begin{equation}
D(\vec{\rho})=\mathcal{L}(\alpha, 2^{2/\alpha}{\cdot}R^2; \vec{\rho}) = \frac{1}{(2\pi)^3} 
\int d^3\vec{\zeta} \, e^{i \vec{\zeta} 
\cdot \vec{\rho}} \, e^{-\left| \vec{\zeta}^{\,T} R^2 \vec{\zeta} \right|^{\alpha/2}},
\end{equation} 
where $\vec \zeta$ is the integration variable, the superscript $T$ denotes the transpose, while $R^2 = \mathrm{diag}\left(R_{\mathrm{out}}^2, R_{\mathrm{side}}^2, R_{\mathrm{long}}^2\right)$ denotes the diagonal matrix of Lévy-scale parameters defined in the Bertsch--Pratt ``out-side-long'' coordinate system~\cite{Pratt1986, Bertsch1988}, neglecting cross-terms. Here the ``long'' direction is the direction of the beam, ``out'' is defined by the average pair momentum, and ``side'' is perpendicular to the previous two. The $2^{2/\alpha}$ factor appears due to $D(\vec{\rho})$ being defined as the autoconvolution of the Lévy-distributed single-particle source.

In this study, for each energy, a total of 4000 minimum-bias events were generated using EPOS4 with its default X3FF Equation of State with cross-over transition and three flavour conservation~\cite{Stefaniak:2022pxc,PhysRevC.89.064903}. 
Out of these, the most central $0-10\%$ were selected via the predefined impact parameter limits given in EPOS4 (a multiplicity-based centrality selection was investigated as part of the systematic uncertainty estimates).
The analysis was performed for identical pion pairs ($\pi^{+}\pi^{+}$ and $\pi^{-}\pi^{-}$). Tracks were selected with $0.15 < p_{T} < 1.0$ GeV/$c$ and $|\eta| < 1$, following typical STAR kinematic settings.
Unlike Ref.~\cite{KORODI2023138295}, where the angle-averaged distance distribution $D(\rho) = \frac{1}{4\pi} \int D(\vec{\rho}) \, d\Omega$ was measured, but similarly to Refs.~\cite{kincses20253DEPOS3,Kincses:2024lnv,Kincses:2025iaf}, we investigate the 3D pair distance distribution.
To express the source coordinates in the Bertsch-Pratt coordinates of the longitudinally co-moving system (LCMS)~\cite{lcms2018Phenix, Pratt1990lcms, lcms2004Phenix}, the components of $\vec{\rho}$ are computed from the particle positions in the laboratory frame as follows:~\cite{Kincses:2024lnv}

\begin{align}
\rho_{\mathrm{out}}^{\mathrm{LCMS}} &= r_x \cos\varphi + r_y \sin\varphi 
- \frac{k_T}{K_0^2 - K_z^2} \left( K_0 t - K_z r_z \right), \\
\rho_{\mathrm{side}}^{\mathrm{LCMS}} &= - r_x \sin\varphi + r_y \cos\varphi, \\
\rho_{\mathrm{long}}^{\mathrm{LCMS}} &= \frac{K_0 r_z - K_z t}{\sqrt{K_0^2 - K_z^2}}.
\end{align}

Here, the azimuthal angle $\varphi$ is defined by $\cos\varphi = {K_x}/{k_T}$, $K_0$ and  $(K_x, K_y, K_z)$ denote the temporal and spatial components of the pair's average four-momentum $\vec{K}$, $k_T = \sqrt{K_x^2 + K_y^2}$ is the pair transverse momentum, and $(r_x, r_y, r_z, t)$ are the particle coordinates in the laboratory frame. Note furthermore that in this paper we calculate $\vec{\rho}$ distributions in distinct bins of transverse mass $m_T=\sqrt{m^2+k_T^2}$, where $m$ is the mass of the investigated particle, in our case that of charged pions. Detailed derivations and conventions can be found in the Methods sections of Refs.~\cite{Kincses:2024lnv, kincses20253DEPOS3}.

For each event, the distribution of relative distances $D(\vec{\rho})$ between pion pairs was constructed in the longitudinally co-moving system. The distance distributions were then fitted using the Lévy parameterization, with the fitting procedure following the methodology outlined in Ref.~\cite{Kincses:2024lnv}, to extract the parameters $\alpha$, $R_{\rm{out}}$, $R_{\rm{side}}$, $R_{\rm{long}}$, and $\lambda$. 
The fit quality was assessed through comparison of one-dimensional projections of theoretical fit curves and simulated distributions (similarly to Fig.5 of Ref.~\cite{Kincses:2024lnv}, and Fig.2 of Ref.~\cite{kincses20253DEPOS3}), ensuring that the main features of the source distributions are captured. The extracted parameters were also checked for consistency with the Lévy walk model and the expected core-core source behavior. The fitted distribution corresponds to the core-core source, consisting of pairs of pions produced close to the interaction point, i.e., primordial pions and the decay products of the shortest lived resonances ~\cite{kincses20253DEPOS3}. In addition, standard fit statistics $\chi^2$/ndf and confidence levels were evaluated to quantify the goodness of fit. Overall, the Lévy parameterization provides a satisfactory description of the freeze-out source across studied transverse mass bins and collision energies, supporting the interpretation of the extracted parameters.

\subsection{Systematic uncertainties}

The systematic uncertainties were estimated by varying several analysis conditions relative to the default configuration. 
The corresponding cuts applied for different sources are summarized for each setting below.

\textbf{(1) Number of averaged events.} Event-by-event studies show that Lévy-shaped sources already appear in individual EPOS events~\cite{KORODI2023138295,Kovacs2025}, indicating that the non-Gaussian features are not caused by event averaging. However, due to limited pair statistics in single events, events were merged in the present analysis to obtain stable Lévy parameter extraction.
It was carefully investigated and confirmed that the extracted parameters converge at a sufficiently large event number. For different collision energies, the number of merged events ($N_{\rm events}$) was varied as follows (the middle value is taken as default):
\begin{itemize}
    \item 7.7, 9.2, 11.5, and 14.5~GeV: (150, 170, 190)
    \item 19.6, 27, 39, and 62.4~GeV: (120, 150, 180)
    \item 130 and 200~GeV: (70, 100, 130)
\end{itemize}

\textbf{(2) Pair momentum difference.} In quantum-statistical correlations, the signal appears at limited values of relative momentum \( Q_{\mathrm{LCMS}} \). Based on the observed $\sqrt{m_T}$ scaling of femtoscopic correlation widths ~\cite{lcms2018Phenix,universe3040085}, we utilize pairs up to a maximal momentum difference of $Q_{\mathrm{LCMS}}^{\mathrm{max}} = c \cdot \sqrt{m_T\cdot 0.001\text{ GeV}}$ where the coefficient \( c \) was varied for different collision energies (again the middle value is the default): 
\begin{itemize}
    \item 7.7, 9.2, 11.5, and 14.5~GeV: (10, 12, 15)
    \item 19.6, 27, 39, 62.4, 130, and 200~GeV: (7, 10, 12)
\end{itemize}

\textbf{(3) Fit range in \(\rho\).} The stability of the extracted source parameters was examined by varying the fitting range in \(\rho\):
\begin{itemize}
    \item 7.7, 9.2, 11.5, and 14.5~GeV: (1-15, 1-20, 1-30)
    \item 19.6, 27, 39, 62.4, 130, and 200~GeV: (1-20, 1-30, 1-40)
\end{itemize}

\textbf{(4) Centrality definition.} Centrality was defined using either the impact parameter ($bim$) from the EPOS4 definition or the charged-particle multiplicity ($ N_{\mathrm{ch}}$). 
The two-dimensional ($bim, N_{\mathrm{ch}}$) distributions of the dataset were analyzed to define centrality bins according to the accumulated event fractions (0-10\%, 10-20\%, etc.). 
The centrality boundaries extracted from the impact parameter distribution are further found to be consistent with those provided directly by EPOS4. Alternatively, we also considered a multiplicity-based event selection. Overall, the choice of centrality definition contributes less than 2\% to the systematic uncertainty of all extracted femtoscopic quantities.

A systematic uncertainty study considering the above points was performed for all collision energies and each \(m_T\) bin. 
The results show that, in the region \(m_T < 0.7~\mathrm{GeV}/c^2\), the systematic uncertainties of the parameters exhibit no significant dependence on collision energy or \(m_T\). In this region, the uncertainties fluctuate mildly around a similar level, with typical magnitudes of about 0.5\%–3\% when averaged over \(m_T\) bins.
For \(m_T > 0.7~\mathrm{GeV}/c^2\), the uncertainties increase noticeably, reaching up to 13\%, primarily due to the reduced pair statistics at higher transverse momentum. When averaged over all \(m_T\) bins, the total systematic uncertainties decrease with increasing collision energy. Consequently, the physical interpretation of the \(m_T\) dependence in this work focuses primarily on the region \(m_T < 0.7~\mathrm{GeV}/c^2\), where the results are more robust and the uncertainties are better controlled.

To provide a concise yet representative summary, Table~\ref{tab:sys_uncertainties}~(a) lists the relative systematic uncertainties for three characteristic collision energies (7.7, 19.6, and 200~GeV), corresponding to the low-, intermediate-, and high-energy regimes, at a representative
\( k_T = 0.425\text{--}0.475~\mathrm{GeV}/c \), 
(corresponding to a mean transverse mass $ m_T \approx 0.469$ GeV/$c^2$).
For each energy, the first four rows show the individual contributions from each source, while the Total row provides the combined systematic uncertainties.
The average total relative systematic uncertainties over all studied $m_T$ intervals 
($0.245$--$0.663~\mathrm{GeV}/c^2$) for the analyzed energies (7.7--200~GeV) are summarized in 
Table~\ref{tab:sys_uncertainties}~(b).

\begin{table}[htbp]
\centering
\caption{Relative systematic uncertainties of Lévy parameters. (a) Relative systematic uncertainties from each source at selected energies: 7.7, 19.6, 200~GeV. (b) Summary of total relative systematic uncertainties over all analyzed energies. }
\label{tab:sys_uncertainties}

\begin{subtable}[t]{\linewidth}
\centering
\caption{Relative systematic uncertainties of the parameters $\alpha$, $\lambda$, $R_{\mathrm out}$, $R_{\mathrm side}$, and $R_{\mathrm long}$ from different sources at three collision energies: 7.7, 19.6, and 200~GeV, for \( k_T = 0.425\text{--}0.475~\mathrm{GeV}/c \),  (corresponding to $m_T \approx 0.470$ GeV/$c^2$).}
\label{tab:sys_uncertainties_sub1}
\footnotesize
\begin{tabular}{l l c c c c c}
\toprule
$\sqrt{s_{_{NN}}}$ &  Source & $\alpha$  & $\lambda$  & $R_\mathrm{out}$  & $R_\mathrm{side}$  & $R_\mathrm{long}$   \\
\toprule
7.7    & $N_{\rm events}$               &  0.6\% &  0.3\% &  0.2\%  &  0.3\%  &  0.5\%  \\   
GeV           & $Q_{\mathrm{LCMS}}^{\rm max}$  &  0.7\% &  0.1\% &  0.2\%  &  1.0\%  &  2.2\%  \\
           & $\rho_{\rm fit}^{\rm min/max}$ &  0.4\% &  0.2\% &  1.0\%  &  0.1\%  &  0.7\%  \\
           & Centrality                     &  0.9\% &  0.2\% &  0.1\%  &  0.6\%  &  0.8\%  \\
           & Total                          &  1.3\% &  0.4\% &  1.1\%  &  1.2\%  &  2.4\%  \\
\midrule
19.6     & $N_{\rm events}$               &  0.3\% &  0.1\% &  0.01\%  &  0.1\%  &  0.2\%  \\   
GeV           & $Q_{\mathrm{LCMS}}^{\rm max}$  &  0.5\% &  0.03\% &  0.4\%  &  1.0\%  &  1.7\%  \\
           & $\rho_{\rm fit}^{\rm min/max}$ &  1.3\% &  0.34\% &  0.4\%  &  0.1\%  &  0.8\%  \\
           & Centrality                     &  0.5\% &  0.2\% &  0.4\%  &  0.5\%  &  0.2\%  \\
           & Total                          &  1.5\% &  0.5\% &  0.7\%  &  1.1\%  &  2.0\%  \\
\midrule
200.0   & $N_{\rm events}$               &  0.1\% &  0.1\% &  0.1\%  &  0.1\%  &  0.1\%  \\   
GeV           & $Q_{\mathrm{LCMS}}^{\rm max}$  &  0.4\% &  0.4\% &  0.2\%  &  1.0\%  &  2.0\%  \\
           & $\rho_{\rm fit}^{\rm min/max}$ &  0.5\% &  0.1\% &  1.0\%  &  0.3\%  &  0.9\%  \\
           & Centrality                     &  0.2\% &  0.2\% &  0.6\%  &  0.4\%  &  0.3\%  \\
           & Total                          &  0.7\% &  0.5\% &  1.1\%  &  1.1\%  &  2.2\%  \\
\bottomrule
\end{tabular}
\end{subtable}

\vspace{1em}

\begin{subtable}[t]{\linewidth}
\centering
\caption{Summary of the average total relative systematic uncertainties (in \%) of the Lévy parameters over all studied $m_T$ intervals ($m_T \approx 0.245\text{--}0.663 ~\mathrm{GeV}/c^2$), for the analyzed $\sqrt{s_{NN}}=7.7\text{--}200$ ~GeV energy range.}
\label{tab:sys_uncertainties_sub2}
\footnotesize
\resizebox{\linewidth}{!}{
\begin{tabular}{c c c c c c c c c c c}
\toprule
$\sqrt{s_{NN}}$ (GeV) & 7.7 &9.2 & 11.5 & 14.5 & 19.6 & 27 & 39 & 62.4 &130 & 200 \\
\midrule
$\alpha$ & 1.1 & 1.3 & 1.2 & 1.1 & 1.5 & 1.5 & 1.2 & 1.24& 1.3 & 1.0 \\
$\lambda$ & 0.6 & 0.6 & 0.5 & 0.6 & 0.7 & 0.5 & 0.6 & 0.6 & 0.5 & 0.5 \\
$R_\mathrm{out}$ & 1.2 & 1.0 & 0.9 & 1.1 & 1.0 & 0.9 & 0.9 & 1.0 & 0.9 & 1.1 \\
$R_\mathrm{side}$ &1.2 & 1.1 & 1.1 & 1.2 & 1.0 & 0.9 & 1.0 & 1.0 & 1.0 & 1.0 \\
$R_\mathrm{long}$ &2.6 & 2.6 & 2.4 & 2.3 & 2.1 & 2.2 & 2.0 & 2.2 & 2.2 & 2.1 \\
\bottomrule
\end{tabular}}
\end{subtable}

\end{table}

\section{Results and discussions}
In this section we present the results for the femtoscopic source parameters as functions of collision energy, focusing on three main aspects: source shape ($\alpha$), source size ($R_{\text{out}},R_{\text{side}},R_{\text{long}}$), and correlation strength ($\lambda$).

\subsection{Source shape}

The Lévy index $\alpha$ characterizes the shape of the emission source:  $\alpha = 2$ corresponds to a Gaussian source, while smaller values indicate long-tailed Lévy-stable distributions, often associated with contributions from long-lived resonances or anomalous diffusion-like dynamics. 

\begin{figure}
    \centering
    \includegraphics[width=\linewidth]{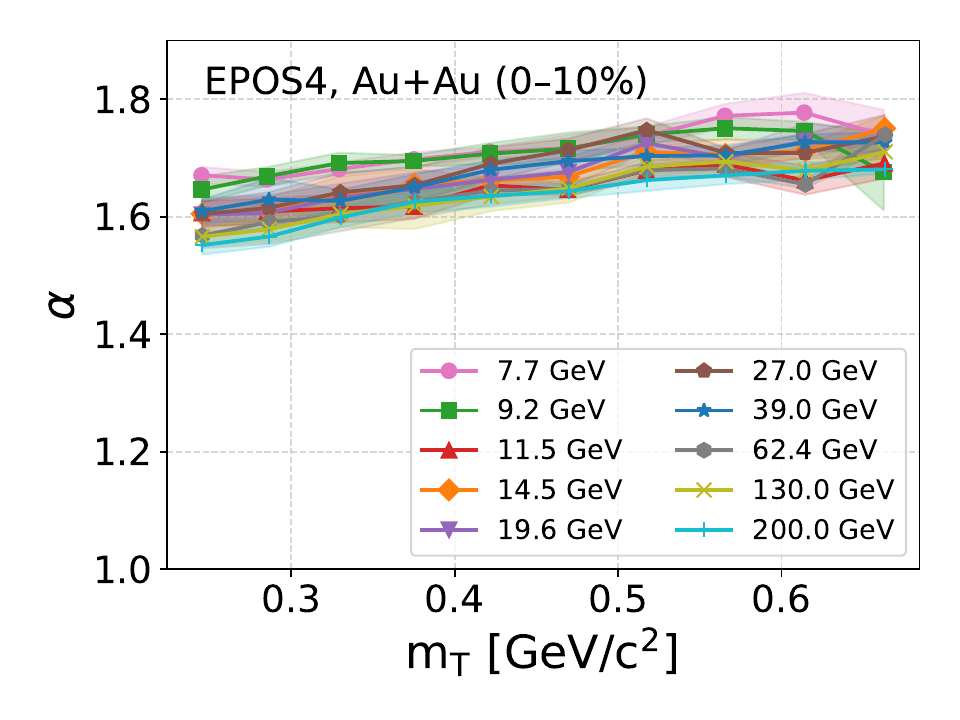}
    \caption{Lévy index $\alpha$ from EPOS4 simulations of $0-10\%$ Au+Au collisions as a function of transverse mass $m_T$ at different energies. The bands show the systematic uncertainties.}
    \label{fig:alpha_mT}
\end{figure}

\begin{figure}
  \centering
    \includegraphics[width=\linewidth]{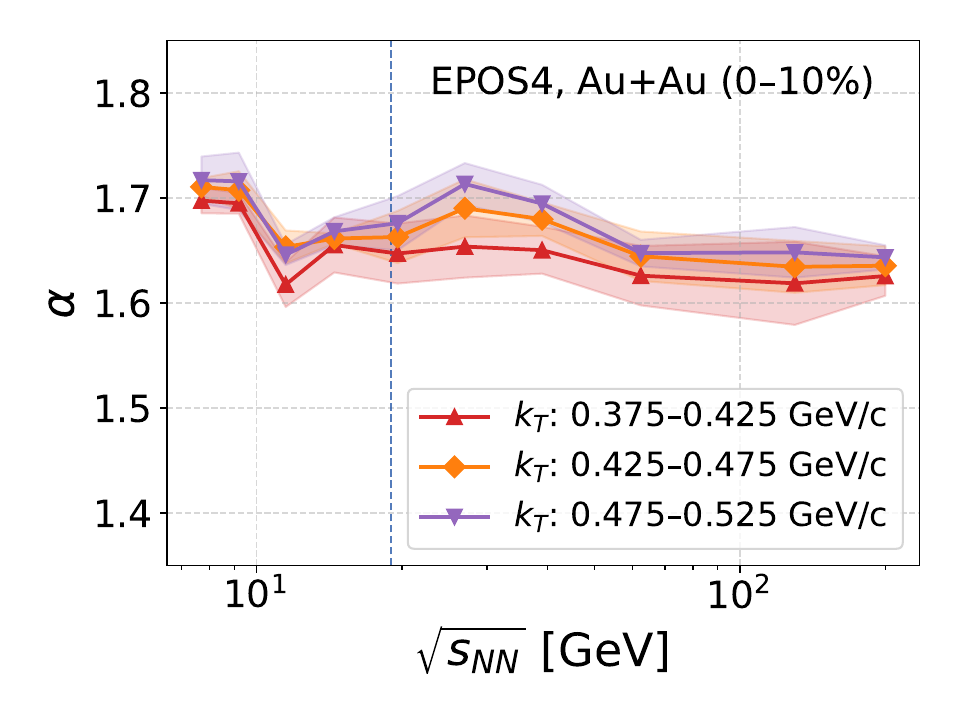}
    \caption{Lévy index $\alpha$ from EPOS4 simulations of $0-10\%$ Au+Au collisions as a function of collision energy $\sqrt{s_{NN}}$ in three bins of $k_T$. The bands show the systematic uncertainties. The vertical blue dashed line at 19~GeV denotes the low-energy region where EPOS4 is expected to be less reliable~\cite{PhysRevC.111.014903}.}
    \label{fig:alpha_Ecm}
\end{figure}

\Fig{fig:alpha_mT} shows the $m_T$ dependence of $\alpha$ for all collision energies. 
Across all collision energies, $\alpha$ exhibits a mild increase with $m_T$, consistent with a reduced relative contribution from long-lived resonances at higher transverse momentum. The trend appears slightly stronger at high energies; however, the slope differences between energies are not statistically significant. \Fig{fig:alpha_Ecm} presents the energy dependence of $\alpha$ for the intermediate $k_T$ region ($0.375 < k_T < 0.525$~GeV/$c$), including three $k_T$ bins: $0.375$–$0.425$, $0.425$–$0.475$~GeV/$c$ and $0.475$–$0.525$  (corresponding to mean $m_T$ values of 0.422, 0.469, and 0.517 GeV/$c^2$, respectively). A localized decrease is observed around $\sqrt{s_{NN}} = 11.5$~GeV, but no clear non-monotonic behavior is present across the energy range, and the overall trend remains nearly energy independent. We note that Ref.~\cite{PhysRevC.111.014903} indicates that EPOS4 is less reliable in the $\sqrt{s_{NN}} < 19.6$~GeV region. To highlight this, a vertical line at 19~GeV is drawn in the figure, to indicate the applicability of EPOS4. For the subsequent energy-dependent plots of the radii (\Fig{fig:radii_Ecm}, \Fig{fig:Rdiff_Ecm}, \Fig{fig:Rratio_Ecm}) and $\lambda$ (\Fig{fig:lambda_Ecm}), the same line is drawn. Given the predicted shape change in femtoscopic sources near the critical point~\cite{Csorgo:2005it}, analyzing this $\alpha(\sqrt{s_{NN}})$ trend for various equations of state is one of the important future research directions.

\subsection{Source size}
The extracted source radii exhibit the expected decrease with increasing average transverse momentum of the pair and an overall growth with increasing collision energy, reflecting the expansion dynamics of the system. The energy dependence of $R_\mathrm{out}$, $R_\mathrm{side}$, and $R_\mathrm{long}$ provides a systematic baseline for comparison with future experimental femtoscopy results.

\begin{figure*}
  \centering
  \includegraphics[width=0.95\linewidth]{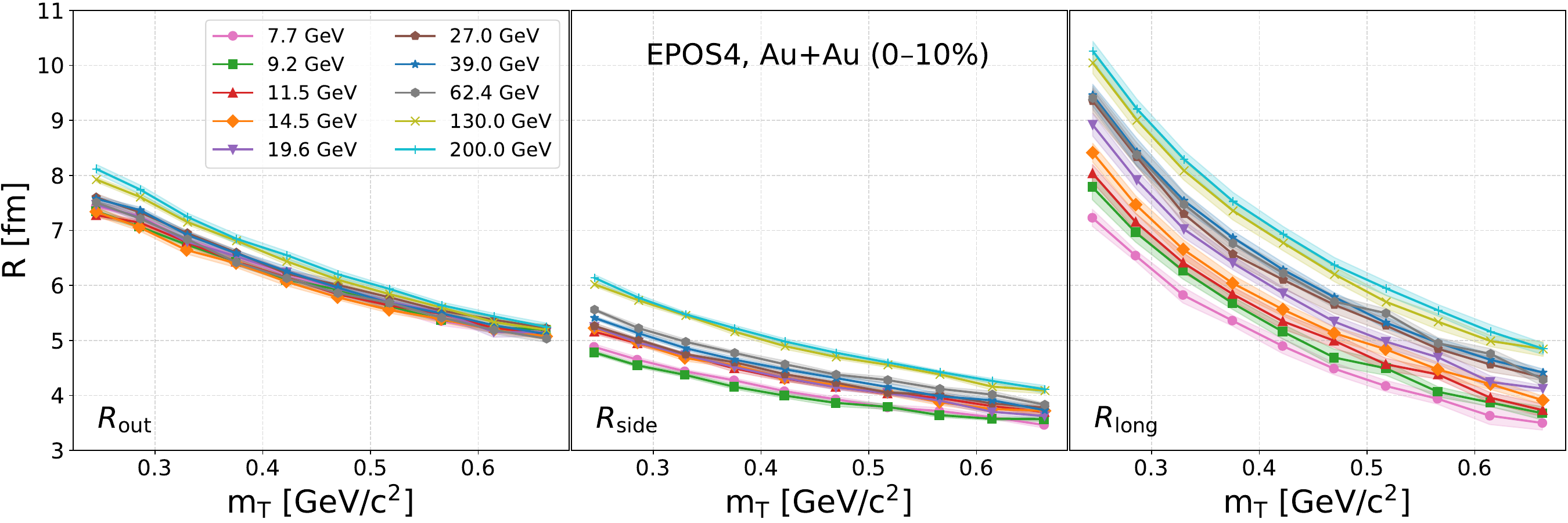}
  \caption{Lévy radii from EPOS4 simulations of $0-10\%$ Au+Au collisions as a function of transverse mass $m_T$ at ${\sqrt{s_{NN}}=7.7-200~\rm{GeV}}$. The bands show the systematic uncertainties.}
  \label{fig:radii_mT}
\end{figure*}

\begin{figure}
  \centering
  \includegraphics[width=0.98\linewidth]{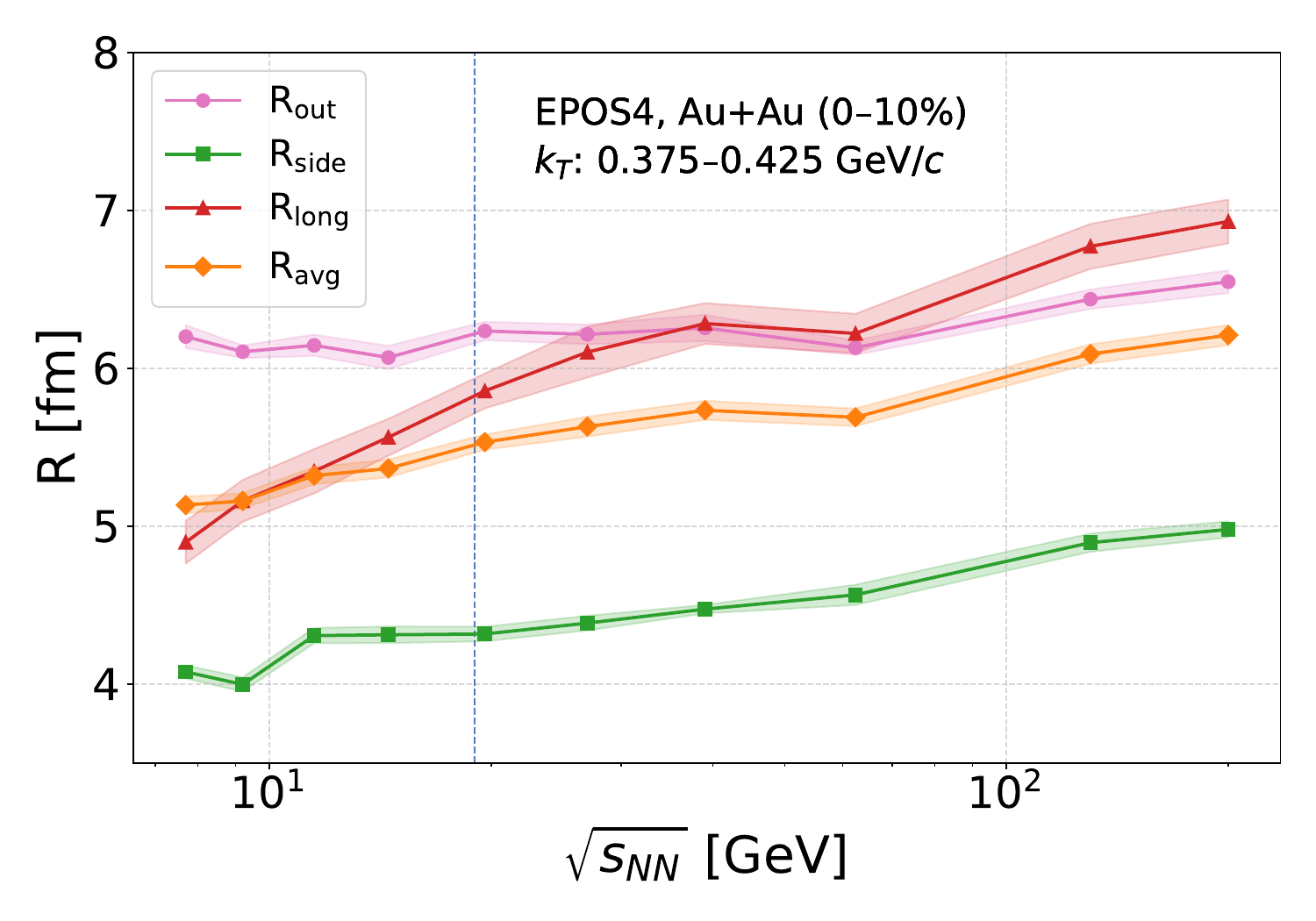}
  \caption{Energy dependence of Lévy radii from EPOS4 simulations of $0-10\%$ Au+Au collisions at ${k_T = 0.375-0.425~\rm{GeV}/c}$. The bands show the systematic uncertainties. The vertical blue dashed line at 19~GeV denotes the low-energy region where EPOS4 is expected to be less reliable~\cite{PhysRevC.111.014903}.}
  \label{fig:radii_Ecm}
\end{figure}

As shown in \Fig{fig:radii_mT}, the femtoscopic radii $R_\mathrm{out}$, $R_\mathrm{side}$, and $R_\mathrm{long}$ exhibit a clear decrease with increasing $m_T$. 
In addition, \Fig{fig:radii_Ecm} presents their energy dependence for the $k_T$ bin of ${0.375-0.425~\rm{GeV}/c}$ (corresponding to ${m_T \approx 0.422~\rm{GeV}/c^2}$), including the averaged radius 
${R_\mathrm{avg} = \sqrt{(R_\mathrm{out}^2 + R_\mathrm{side}^2 + R_\mathrm{long}^2)/3}}$. 
It can be seen that $R_\mathrm{long}$ and $R_\mathrm{side}$ increase with collision energy. 
The increase of $R_\mathrm{long}$ is consistent with a longer system lifetime at higher collision energies. 
In EPOS4, this originates from the higher initial energy density and the correspondingly longer hydrodynamic evolution. In contrast, $R_\mathrm{out}$ exhibits a rather weak dependence on energy. This behavior was already observed in the STAR data ~\cite{STAR:2014shf}.
The gradual increase of $R_\mathrm{long}$ and $R_\mathrm{side}$ with collision energy, reflecting spatial expansion at higher energies, is further illustrated by the pair distance distributions shown in \Fig{fig:D_rho}. 
We show $D(\rho_\mathrm{out}^\mathrm{LCMS})$, $D(\rho_\mathrm{side}^\mathrm{LCMS})$, and $D(\rho_\mathrm{long}^\mathrm{LCMS})$ for the $k_T$ bin $0.225-0.275~\rm{GeV}/c$, corresponding to ${m_T \approx 0.285~\rm{GeV}/c^2}$), at low (7.7 GeV), intermediate (19.6 GeV), and high (200 GeV) energies. A clear widening of $\rho_\mathrm{long}$ and $\rho_\mathrm{side}$ distributions is observed, indicating an increase of the longitudinal and transverse source sizes with energy, while the $\rho_\mathrm{out}$ distributions show no significant change, consistent with the weak energy dependence of $R_{out}$ discussed above.

\begin{figure}
  \centering
  \includegraphics[width=0.98\linewidth]{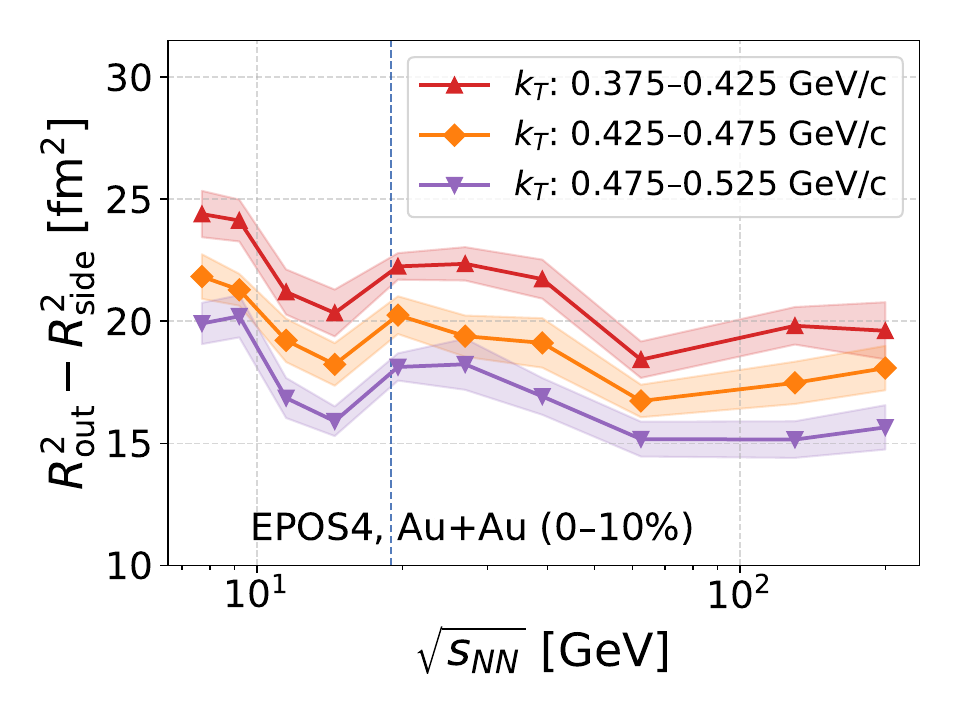}                  
    \caption{The radius difference $R_\mathrm{diff}^2$ from EPOS4 simulations of $0-10\%$ Au+Au collisions as a function of collision energy $\sqrt{s_{NN}}$ in three $k_T$ bins. The bands show the systematic uncertainties. The vertical blue dashed line at 19~GeV denotes the low-energy region where EPOS4 is expected to be less reliable~\cite{PhysRevC.111.014903}.}
    \label{fig:Rdiff_Ecm}
\end{figure}

\begin{figure}
  \centering
  \includegraphics[width=0.98\linewidth]{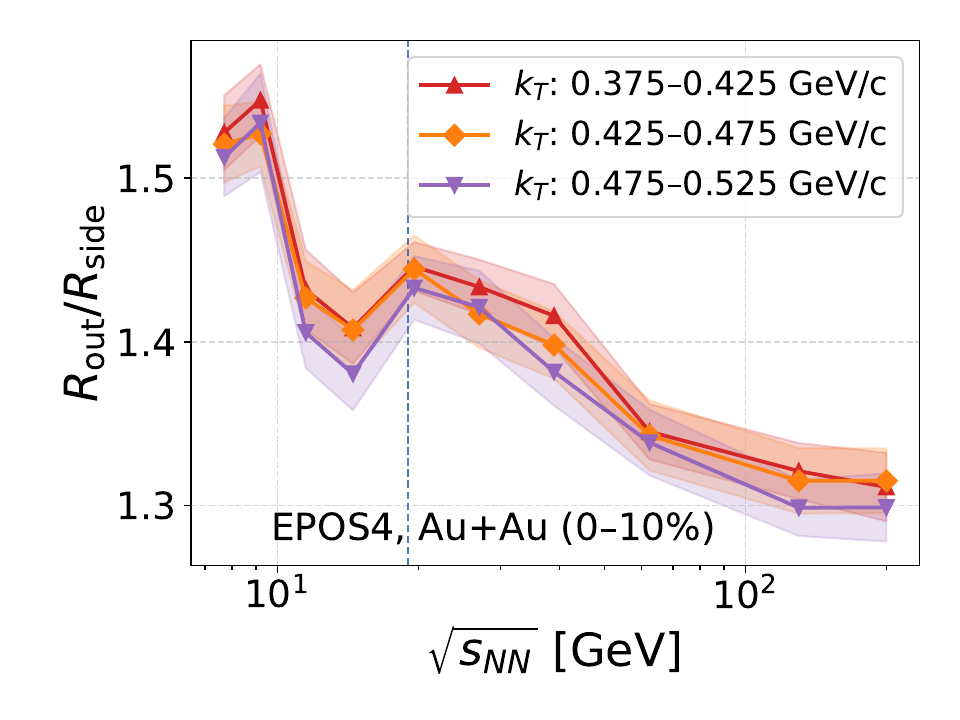}                  
    \caption{The radius ratio $R_\mathrm{out}/R_\mathrm{side}$ from EPOS4 simulations of $0-10\%$ Au+Au collisions as a function of collision energy $\sqrt{s_{NN}}$ at ${k_T=0.375-0.525~\rm{GeV}/c}$. The bands show the systematic uncertainties. The vertical blue dashed line at 19~GeV denotes the low-energy region where EPOS4 is expected to be less reliable~\cite{PhysRevC.111.014903}.}
    \label{fig:Rratio_Ecm}
\end{figure}

Ref.~\cite{Csanád_2025} has highlighted the sensitivity of the Lévy scale difference $R_{\mathrm{diff}}^2 = R_{\mathrm{out}}^{2} - R_{\mathrm{side}}^{2}$ to potential critical-point 
signatures in the QCD phase diagram. Motivated by this, we investigated the energy dependence of both $R_{\mathrm{diff}}$ and the ratio $R_{\mathrm{out}}/R_{\mathrm{side}}$, as shown in \Fig{fig:Rdiff_Ecm} and \Fig{fig:Rratio_Ecm}.
The quantity $R_{\mathrm{diff}}$ decreases with increasing $m_T$, consistent with the behavior of the individual femtoscopic radii. And its collision-energy dependence shows 
a non-monotonic pattern, exhibiting a noticeable drop at 14.5~GeV followed by partial recovery and a gradual decrease, yielding no clear global trend across the full energy range.
The ratio $R_{\mathrm{out}}/R_{\mathrm{side}}$, on the other hand, exhibits a stronger overall decrease with increasing collision energy, accompanied by a notable pronounced enhancement in the interval $14.5 - 39$~GeV.
The strong deviation of this ratio from unity, especially at the lowest energies, is probably stemming from a not fully adequate description of several details, such as pre-thermalized acceleration, equation of state, or viscous corrections~\cite{Pratt:2008qv}. The investigation of these observables for different equations of state is of crucial importance and presents the focus of subsequent studies.

\subsection{Correlation strength}

The correlation strength parameter $\lambda$ is sensitive~\cite{Csorgo:1994in,Bolz:1992hc} to the fraction of coherent emission, long-lived resonance contributions, and purity corrections. In simulations, $\lambda$ can be extracted from the integral of the fitted core-core component of the source~\cite{kincses20253DEPOS3}. 
Values extracted in this EPOS4 analysis show a moderate decrease with increasing collision energy, which may be attributed to the enhanced role of resonance decays at higher energies. 

\begin{figure}
  \centering
    \includegraphics[width=0.98\linewidth]{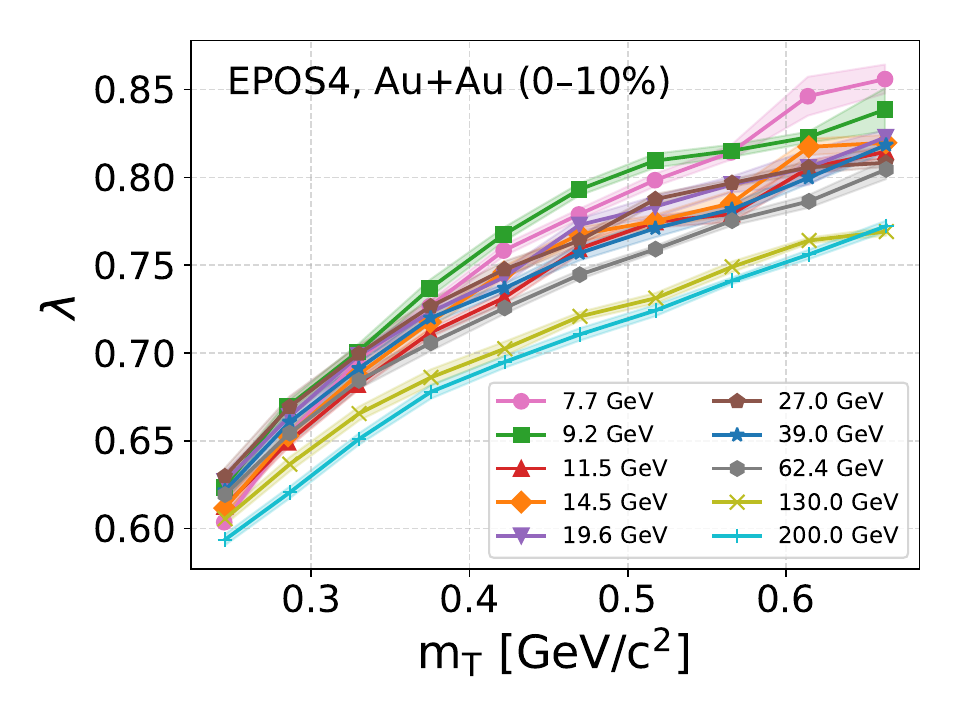}
    \caption{Correlation strength parameters $\lambda$ from EPOS4 simulations of $0-10\%$ Au+Au collisions as a function of transverse mass $m_T$ at $\sqrt{s_{NN}}=7.7-200~\rm{GeV}$. The bands show the systematic uncertainties.}
    \label{fig:lambda_mT}
\end{figure}

\begin{figure}
  \centering
    \includegraphics[width=0.98\linewidth]{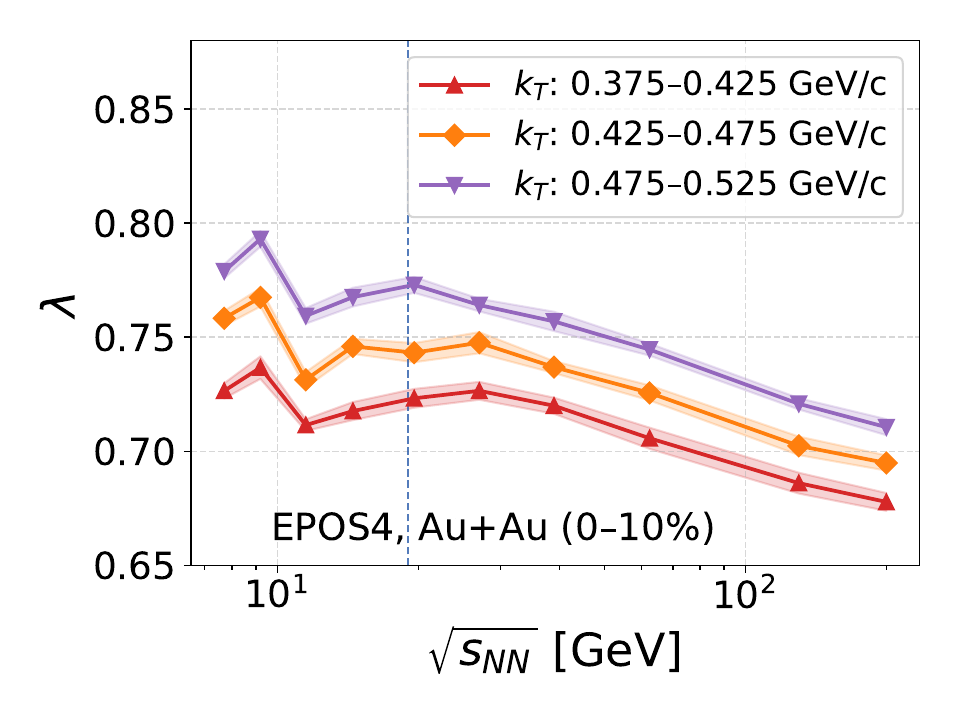}
    \caption{Energy dependence of $\lambda$ from EPOS4 simulations of $0-10\%$ Au+Au collisions at $k_T = 0.375-0.525~\rm{GeV}/c$. The bands show the systematic uncertainties. The vertical blue dashed line at 19~GeV denotes the low-energy region where EPOS4 is expected to be less reliable~\cite{PhysRevC.111.014903}.}
    \label{fig:lambda_Ecm}
\end{figure}

The transverse mass dependence of the extracted $\lambda$ parameter at various collision energies is presented in \Fig{fig:lambda_mT}, showing an overall increase with $m_T$. 
As shown in \Fig{fig:lambda_Ecm}, $\lambda$ reveals a decreasing trend with increasing collision energy, although noticeable anomalies are observed at $\sqrt{s_{NN}} = 9.2$ and $11.5$~GeV, where the values deviate from the general behavior.

\subsection{Comparison with EPOS3 and experimental data}

In this analysis, the femtoscopic parameters of pions from $200$~GeV Au+Au collisions are generally consistent with those obtained from EPOS3 in Ref.~\cite{kincses20253DEPOS3} within approximately $2\sigma$, with the notable exception of the $R_{\mathrm{side}}$ radius, which exhibits a systematic reduction.
For $R_{\mathrm{side}}$, EPOS4 predicts systematically smaller values, with deviations corresponding to ${N_{\sigma} = 2.5-5.8}$, where
$N_{\sigma} = {\Delta}/{\sigma_{\mathrm{combined}}}$
and $\Delta$ is the difference between the EPOS4 and EPOS3 results, while 
${\sigma_{\mathrm{combined}} = \sqrt{\sigma_{\mathrm{EPOS4}}^{2} + \sigma_{\mathrm{EPOS3}}^{2}}}$
denotes the quadrature sum of their uncertainties. 
The $N_{\sigma}$ deviations between EPOS4 and EPOS3 for each Lévy parameter, evaluated in different $m_{T}$ bins, are listed in Table~\ref{tab:Nsigma}. Due to this similarity to EPOS3, and the detailed data comparison of EPOS3 to data of the one-dimensional measurement~\cite{Abdulameer2024RHIC} in Ref.~\cite{kincses20253DEPOS3}, it can be stated that EPOS4 is also compatible with the data. However, it will be important to compare our predictions with three-dimensional measurements, once these are performed.

\begin{table}
\centering
\caption{$N_{\sigma}$ significance of the differences between EPOS4 and EPOS3 (EPOS4 $-$ EPOS3) for the Lévy parameters at different transverse mass in Au+Au collisions at $\sqrt{s_{NN}}$ = 200 GeV.}
\vspace{0.5em}
\label{tab:Nsigma}
\resizebox{\linewidth}{!}{
\begin{tabular}{c l l l l l l}
\hline
$m_{T}$ (GeV/$c^{2}$) 
& $N_\sigma(\alpha)$
& $N_\sigma(\lambda)$
& $N_\sigma(R_{\text{out}})$
& $N_\sigma(R_{\text{side}})$
& $N_\sigma(R_{\text{long}})$
& $N_\sigma(R_{\text{avg}})$ \\
\hline
0.24 & 1.8 &           -0.095  & \phantom{-}0.11 & -5.8 & -1.3 & -2.3 \\
0.28 & 1.3 &           -0.12   &           -0.37 & -5.3 & -0.96 & -1.6 \\
0.33 & 2.1 &           -0.084  &           -0.25 & -4.7 & -0.74 & -1.3 \\
0.38 & 2.3 &           -0.044  & \phantom{-}0.34 & -3.8 & -0.66 & -1.1 \\
0.42 & 1.8 &           -0.088  & \phantom{-}1.2  & -3.6 & -0.39 & -0.75 \\
0.47 & 1.0 &           -0.088  & \phantom{-}1.5  & -3.1 & -0.31 & -0.64 \\
0.52 & 1.5 &           -0.15   & \phantom{-}1.8  & -2.8 & -0.25 & -0.57 \\
0.57 & 0.20 &          -0.056  & \phantom{-}1.7  & -2.8 & -0.068 & -0.49 \\
0.62 & 0.70 &          -0.061  & \phantom{-}2.0  & -2.6 & -0.23 & -0.54 \\
0.66 & 0.12 & \phantom{-}0.031 & \phantom{-}2.8  & -2.5 & -0.26 & -0.52 \\
0.71 & 0.68 & \phantom{-}0.025 & \phantom{-}3.6  & -2.5 & -0.36 & -0.58 \\
\end{tabular}
}
\end{table}

\section{Summary and outlook}
We present a detailed investigations of the simulated particle emitting source in 0-10\% central Au+Au collisions at collision energies $\sqrt{s_{NN}}=7.7-200$ GeV within the framework of the EPOS4 model. The spatial pair distance distribution $D(\vec \rho)$ for identical charged pions was calculated in the longitudinally comoving system for several bins of pair transverse mass, and its one-dimensional projections in the Bertsch-Pratt out, side, and long coordinates were obtained. These projections were simultaneously fitted with those of a single 3D Lévy distribution, providing a statistically and qualitatively adequate description of the source up to distances of a few dozen (or up to a hundred) femtometers. The following source parameters were extracted: Lévy index (or shape parameter) $\alpha$, spatial Lévy scales (or HBT radii) $R_{\rm out,side,long}$, and correlation strength parameter $\lambda$. And then the transverse mass $m_T$ and energy dependence of these source parameters was investigated.

We found that the radii $R_{\rm out,side,long}$ decrease with $m_T$ at all collision energies, a characteristic observation attributed to the expansion of the system. This decrease appeared to be the strongest in the longitudinal direction. The radii show a moderate increase with collision energy, also strongest for the longitudinal direction. An average scale ($R_{\rm avg}$) and out-side anisotropy (in form of the difference and the ratio of the out and side radii were) also investigated. The anisotropy appears to decrease towards higher collision energies. 
The Lévy exponent, describing the source shape, was also investigated as a function of $m_T$ and collision energy. A weak but smooth increase with $m_T$ is observed across all energies, whereas a less regular decrease occurs with increasing collision energy. 
The correlation strength parameter $\lambda$, extracted based on the integral of the fitted Lévy distribution, and corresponding to the core fraction among all pion pairs, was also investigated as a function of transverse mass and collision energy. A clear increase is found with $m_T$ for all collision energies. This can also be described as a ``hole'' at small $m_T$, as discussed by PHENIX in Ref.~\cite{Abdulameer2024RHIC}. 
In EPOS4, the predicted $\lambda$ values at $\sqrt{s_{NN}}$ = 200 GeV are very close to those obtained in EPOS3~\cite{kincses20253DEPOS3}, which were shown to describe the PHENIX data well.
Furthermore, a weak decrease is found in $\lambda$ for larger collision energies.

One important future research direction is to investigate the equation of state dependence of the Lévy source observables. In particular, a modified equation of state, where the order of the phase transition changes, is expected to leave characteristic signatures in the increase of the correlation radius in the out direction ($R_{\rm out}$). Similarly, the $\alpha$ parameter is expected to be modified due to criticality, with a decrease near the critical point~\cite{Csorgo:2005it}, making it essential to calculate $\alpha$ versus collision energy for an equation of state describing a second-order phase transition, as well as for the case of enhanced non-hydrodynamic modes. Finally, we note that it will be crucial to investigate the equation of state dependence of these observables, and compare them to data, once Lévy source parameter measurements at several collision energies will be available.

\section*{Acknowledgements}
This research was funded by the NKFIH grants TKP2021-NKTA-64, PD-146589, K-146913, K-138136, and NKKP ADVANCED 152097.



\end{document}